\documentclass[3p,times,twocolumn]{elsarticle}

\usepackage{ecrc}

\volume{00}

\firstpage{1}

\journalname{Nuclear Physics B Proceedings Supplement}

\runauth{L. Alvarez-Ruso}

\jid{nuphbp}

\jnltitlelogo{Nuclear Physics B Proceedings Supplement}

\usepackage{amssymb}
\usepackage{amsthm}

\usepackage{amsmath} % mine

\def\be{\begin{equation}}
\def\ee{\end{equation}}
\def\bea{\begin{eqnarray}}
\def\eea{\end{eqnarray}}
\def\bear{\begin{array}}
\def\ear{\end{array}}
\def\bfig{\begin{figure}}
\def\efig{\end{figure}}
\def\bcen{\begin{center}}
\def\ecen{\end{center}}
\def\bi{\begin{itemize}}
\def\ei{\end{itemize}}

\def\raw{\rightarrow}

\def\slash{\!\!\! /}
\begin{document}

\begin{frontmatter}



\dochead{}

\title{Neutrino interactions: challenges in the current theoretical picture}

\author{Luis Alvarez-Ruso}

\address{Centro de F\'{\i}sica Computacional, Departamento de F\'{\i}sica, Universidade de Coimbra, Portugal}

\begin{abstract}
The present theoretical status of neutrino interactions in the few-GeV region is reviewed. Quasielastic scattering, pion production, photon emission and their importance for neutrino oscillation studies are discussed, making emphasis on the open questions that arise in the comparison with new experimental data. 
\end{abstract}


\end{frontmatter}

\section{Introduction and motivation}
\label{intro}

Recent years have witnessed an intense experimental and theoretical activity aimed at a better understanding of neutrino interactions with nucleons and nuclei. While the main motivation for these efforts is the demand from oscillation experiments in their quest for a precise determination of neutrino properties, the relevance of neutrino interactions with matter is more far-reaching. They are important for astrophysics, physics beyond the standard model, hadronic and nuclear physics.    

In the few-GeV neutrino-energy region, where most oscillation experiments operate, the dominant reaction channel through which neutrinos reveal themselves (and their flavor) is charged-current quasielastic scattering (CCQE)  $\nu_l \, n \raw l^- \, p$. Oscillation probabilities depend on the neutrino energy, unknown for broad fluxes and usually obtained from the measured angle and energy of the outgoing lepton using two-body kinematics
\be
\label{enu}
E_\nu = \frac{2 m_n E_l - m_l^2 - m_n^2 + m_p^2}{2 \left( m_n - E_l + \sqrt{E_l^2 -m_l^2} \cos{\theta_l} \right)}  \,.
\ee 
This determination is only exact for free neutrons and under the condition that inelastic events (mainly pion production ones) are identified and excluded. For nuclear targets, as in all modern neutrino experiments,  
the situation is more involved as shown in Fig.~\ref{genie}.  Here the reconstructed neutrino-energy distribution is obtained applying Eq.~(\ref{enu}) to charged-current (CC) events generated with the GENIE Monte Carlo (MC) generator~\cite{Andreopoulos:2009rq} for collisions of a monochromatic 1~GeV neutrino beam on $^{16}$O. 
\bfig[h!]
\bcen
\includegraphics[width=1.08\linewidth]{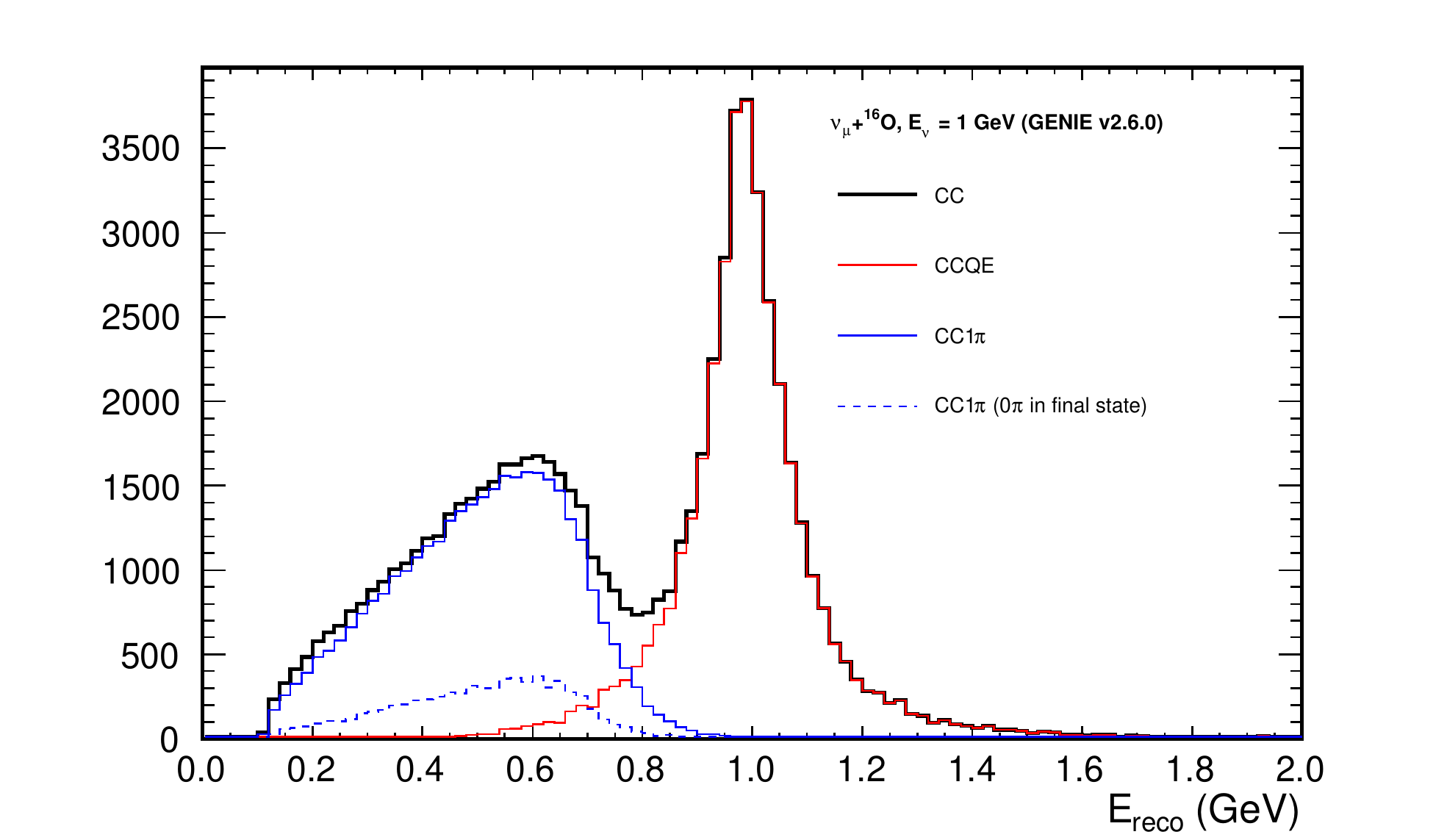}
\caption{(color online) Reconstructed neutrino-energy distribution and its main components for a 1~GeV incident neutrino beam and $^{16}$O target.}
\label{genie}
\ecen
\efig  
Instead of a sharp line at the incident energy, the plot shows a broad peak due to Fermi motion. More importantly, there is a peak of non-CCQE events for which $E_\nu$ reconstruction according to Eq.~(\ref{enu}) is wrong. Many of them can be excluded by detecting the emitted pions, but even with ideally perfect detection efficiency, a remaining (CCQE-like) fraction of events corresponding to absorbed pions introduces a systematic error in the neutrino energy reconstruction that affects the determination of oscillation parameters (see for instance Ref.~\cite{Leitner:2010kp}). Fake CCQE events can only be removed using a model dependent MC simulation. Therefore, a good insight into the dynamics of neutrino-nucleus ($\nu A$) collisions can be hardly underestimated.

The importance of weak pion production ($\pi$P) for oscillations studies is not limited to the contamination of CCQE samples. Neutral current (NC) $\pi^0$ production (both incoherent and coherent) is a large background for $\nu_e$ appearance searches. When one of the two photons from a $\pi^0$ decay escapes detection, the $\pi^0$ cannot be distinguished from an electron born in a $\nu_e$ induced CC interaction. Therefore, a precise determination of $\theta_{13}$ and the potential discovery of CP violation in the lepton sector requires that this background is reliably subtracted. Another $\nu_e$ appearance background is photon emission in NC, one of the relevant processes at
low reconstructed neutrino energies where MiniBooNE has found an unexpected excess of events~\cite{AguilarArevalo:2007it}. 

The experimental information on neutrino interactions is rapidly growing. Detailed quasielastic (QE) scattering measurements have been published by MiniBooNE at $E_{\nu} \sim 1$~GeV (CCQE~\cite{AguilarArevalo:2010zc}, NCQE~\cite{AguilarArevalo:2010cx}) and NOMAD at high (3-100~GeV) energies~\cite{Lyubushkin:2008pe} (CCQE). MiniBooNE has also reported single pion production cross section measurements on different channels: CC$\pi^+$~\cite{AguilarArevalo:2010bm}, CC$\pi^0$~\cite{AguilarArevalo:2010xt} and NC$\pi^0$~\cite{AguilarArevalo:2009ww}, including first single- and double-differential distributions. Special attention has been paid to coherent $\pi$P in both the CC and NC channels by MiniBooNE~\cite{AguilarArevalo:2009ww}, SciBooNE~\cite{Hiraide:2008eu,Kurimoto:2010rc} and NOMAD~\cite{Kullenberg:2009pu}. These new data represent a major improvement and are challenging our understanding of $\nu A$ interaction physics.

\section{Modelling quasielastic scattering}   

At the nucleon level, the weak interaction is defined by the current
\bea
J_\alpha & = & \bar{u}_{N'}  
\left[ \gamma_{\alpha} F_1 + \frac{i}{2m_N} \sigma_{\alpha\beta} \, q^{\beta} F_2  \right. \nonumber \\
& & \left. -\gamma_{\alpha}\gamma_5 F_A - \frac{q_{\alpha}}{m_N} \gamma_5 F_P \right] u_N \,,
\label{curr}
\eea
written in terms of form factors (FF) $F_{1,2,A,P}(Q^2)$. Vector-current conservation and isospin symmetry imply that $F_{1,2}$ are given in terms of the nucleon electromagnetic FF, extracted from electron scattering data~\cite{Bodek:2007ym}. Partial conservation of the axial current (PCAC) allows to relate $F_P$  to $F_A$, usually parametrized in a dipole form
\be
F_A (Q^2) = g_A \left( 1 + \frac{Q^2}{M_A^2} \right)^{-2} 
\ee 
in analogy to the electric FF of the proton at low $Q^2$. Once $g_A$ is fixed from neutron $\beta$ decay, the axial mass $M_A$, related to the axial mean square radius $\langle r_A^2 \rangle = 12/M_A^2$, remains the only unknown nucleon property in Eq.~(\ref{curr}). The value of $M_A$ extracted from early CCQE measurements on deuterium targets is $M_A = 1.016 \pm 0.026$~GeV~\cite{Bodek:2007ym}. While one might be tempted to distrust this result based on experiments with low statistics and poorly know neutrino fluxes, there are good reasons to think that, at least at low $Q^2$, $M_A \sim 1$~GeV. Indeed, there is a low energy theorem that relates $\pi$ electroproduction amplitudes to $F_A$ at threshold and in the chiral limit. Using models to connect this theorem with data it has been found that $\tilde{M}_A = 1.069 \pm 0.016$~GeV~\cite{Liesenfeld:1999mv}. Moreover, applying a hadronic correction that can be precisely calculated at low $Q^2$ using chiral perturbation theory~\cite{Bernard:1992ys}, the resulting  $M_A=1.014 \pm 0.016$~GeV is even closer to the one from $\nu d$ experiments.     

The vast experience acquired in electron-nucleus scattering studies has been applied to the $\nu A$ case. The simplest model in the QE region, present in most event generators used in the analysis of neutrino experiments, is the relativistic global Fermi gas (RFG)~\cite{Smith:1972xh}. It assumes the impulse approximation (IA) according to which 
the interaction takes place on single nucleons whose contributions are summed incoherently. The struck nucleons have momentum distributions characterized by a Fermi momentum $p_F$, and a constant binding energy $\epsilon_B$. Outgoing nucleons cannot go into occupied states (Pauli blocking). Such a simple picture explains qualitatively inclusive QE electron scattering data but fails in the details. A better description requires a more realistic treatment of nuclear dynamics. Interacting  nucleons do not have a well defined dispersion relation but become broad states characterized by spectral functions (SF)
\be
S_{h,p} = - \frac{1}{\pi} \frac{\mathrm{Im}  \Sigma}{(p^2-m_N^2-\mathrm{Re} \Sigma)^2 + (\mathrm{Im} \Sigma)^2} \,,
\ee
where $\Sigma$ is the nucleon selfenergy. The particle SF $S_p$ accounts for the interaction of the outgoing nucleon with the medium. The hole SF $S_h$ includes an 80-90\% contribution from single-particle states while the rest of the nucleons participate in nucleon-nucleon interactions (correlations) and are located at high momentum~\cite{Benhar:2005dj,Ankowski:2007uy}).  In several models these correlations are neglected. An example is the relativistic mean field (RMF) model~\cite{Maieron:2003df} where the initial nucleons are treated as single-particle bound states whose wave functions are solutions of the Dirac equation with a $\sigma$-$\omega$ mean field potential. The same energy independent real potential and Dirac equation are used to obtain the distorted wave function of the outgoing nucleons. The fact that the potential is real makes the model suitable for inclusive processes, in contrast to those with complex optical potentials more appropriate for exclusive reactions like $A(\nu_l,l^- p)$ because the imaginary (absorptive) part of the potential accounts for the flux loss towards other channels (see for instance Figs.~1-2 of Ref.~\cite{Meucci:2004ip}). 

Another alternative to the RFG is the so called Local Fermi Gas (LFG) where the Fermi momentum depends on the coordinate through the nuclear density profile $p_F (r) = [(3/2) \pi^2 \rho(r)]^{1/3}$. The LFG description introduces space-momentum correlations, absent in the GFG, that render more realistic the nucleon momentum distributions (see Fig.~6 of Ref.~\cite{Leitner:2008ue}). A great advantage of LFG is that, owing to its simplicity, microscopic many-body effects such as SF~\cite{Leitner:2008ue,Nieves:2004wx} and long range random phase approximation (RPA) correlations~\cite{Singh:1992dc,Nieves:2004wx,Martini:2009uj} are tractable in a realistic manner. 

\subsection{The CCQE puzzle}

A common feature of all known calculations of the CCQE integrated cross section on $^{12}$C applying the different theoretical techniques outlined above is that they underestimate recent MiniBooNE data. This surprising situation is illustrated in Fig.~\ref{CCQE} for some of the model calculations collected in Ref.~\cite{Boyd:2009zz} and also the RFG. Theoretical results from quite different models lie on a rather narrow band (narrower than the experimental errorbars) clearly below the data: at $E_\nu = 0.8$~GeV, $\sigma_{th} \sim 4.5-5$ while $\sigma_{exp}  \sim 7 \times 10^{-38}$~cm$^2$. The differences in the nucleon FF adopted in these calculations are minor. In particular all take $M_A \sim 1$~GeV.  
\bfig[h!]
\bcen
\includegraphics[width=\linewidth]{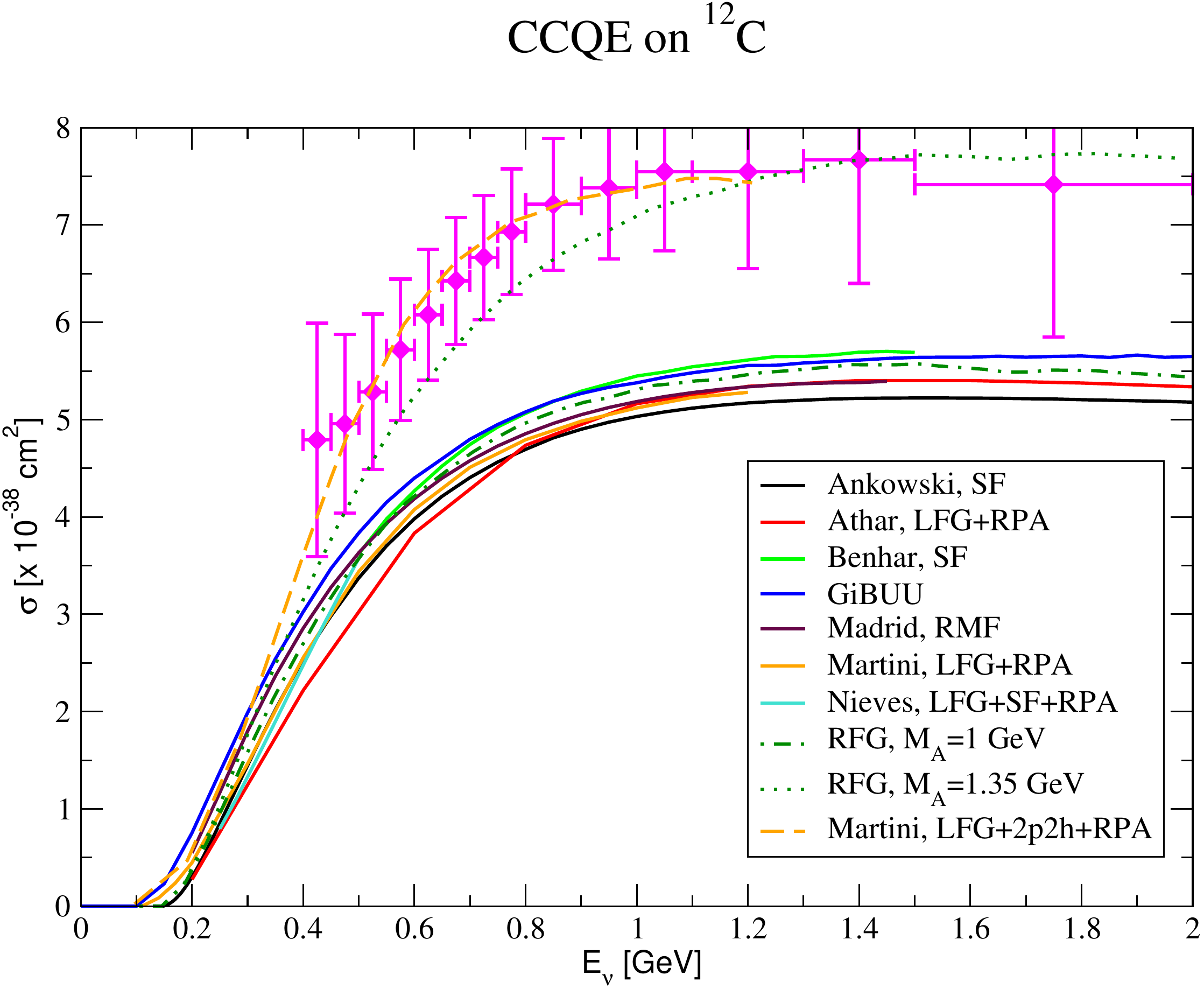}
\caption{(color online) Summary of CCQE total cross sections. Solid lines denote the models from Refs~\cite{Ankowski:2007uy}, \cite{SajjadAthar:2009rd}, \cite{Benhar:2006nr}, \cite{Leitner:2006ww,Leitner:2008ue}, \cite{Maieron:2003df}, \cite{Martini:2009uj} and \cite{Nieves:2004wx} in this order, as reported in Ref.~\cite{Boyd:2009zz}. The dash-dotted and dotted lines are RFG calculations with $p_F = 220$~MeV, $\epsilon_B =34$~MeV and $M_A = 1$ and 1.35~GeV respectively. The dashed line is the result of Ref.~\cite{Martini:2009uj} after adding the $2p-2h$ contributions. The data points are from MiniBooNE~\cite{AguilarArevalo:2010zc}.}
\label{CCQE}
\ecen
\efig      

Several interpretations of this discrepancy are currently under debate. One points at the difficulty that the neutrino-flux determination represents and the possibility that its absolute normalization has been underestimated. On the other hand, according to the MiniBooNE collaboration the systematic errors in the flux estimation have been determined by
varying parameters within their uncertainties and accounting for correlations~\cite{AguilarArevalo:2008yp} so it is legitimate to expect that the errorbars account for the uncertainties in the flux normalization. Another strategy is to extract $M_A$ from MiniBooNE data. In Ref.~\cite{AguilarArevalo:2010zc}, a fit to the shape of the reconstructed $Q^2$ distribution with the RFG model yielded $M_A=1.35 \pm 0.17$~GeV, which is much higher than the world average and the recent NOMAD result $M_A = 1.05 \pm 0.02(stat) \pm 0.06(syst)$~GeV~\cite{Lyubushkin:2008pe} at high energies. The integrated cross section computed with the new value of $M_A$ is consistent with the normalized data as can be seen in Fig.~\ref{CCQE} (dotted line). A similar $Q^2$ fit but using a more elaborated distorted wave IA model, like the RMF sketched above but taking also into account nuclear correlations for the initial nucleons and with a different (real) potential for the outgoing ones, obtained $M_A = 1.37$~GeV~\cite{Butkevich:2010cr}. A better description of the low $Q^2$ region compared to the RFG was also achieved. With a state-of-the-art SF, the best $Q^2$ fit and a good description of muon energy spectrum and angular distribution were found for an $M_A$ as large as 1.6~GeV~\cite{Benhar:2010nx}. With a similar SF but fitting directly the measured flux-averaged double differential cross sections $\langle d^2\sigma /dE_\mu d\cos{\theta_\mu} \rangle$ taking into account the flux uncertainty and introducing a three-momentum cut of 500~MeV to exclude the IA breakdown region, it was obtained that $M_A = 1.343 \pm 0.060$~GeV, lower but still incompatible with earlier determinations.

A third possibility has been put forward by Martini et al.~\cite{Martini:2009uj}. They have studied inclusive $\nu A$ scattering in a LFG using RPA and taking into account two-particle-two-hole ($2p-2h$) contributions, in particular some terms that are not part of the SF (see diagrams 2, 3, 3' in Fig.~1 of Ref.~\cite{Martini:2009uj}).  As shown in Fig.~\ref{CCQE}, with $1p-1h$ excitations alone, the prediction of this model is consistent with the rest but the $2p-2h$ component turns out to be substantially large and capable of explaining the size of the cross section measured by MiniBooNE. An interesting prediction of the model concerns the antineutrino CCQE reaction~\cite{Martini:2010ex}: the different interaction pattern implies that, contrary to the neutrino case, $2p-2h$ excitations play a minor role.      
These suggestions should be further investigated by comparing the $Q^2$ distribution and double-differential cross sections with data. A detailed validation with inclusive electron scattering data is also desirable. The role of meson-exchange currents (MEC) and relativistic effects needs to be elucidated. Work in this direction has already started with the calculation of vector MEC in the $2p-2h$ sector with the RFG~\cite{Amaro:2010sd}.

\section{Modelling incoherent pion production}

Pion production in nuclei is incoherent when the final nucleus is excited ($\nu_l \, A \raw l \, \pi \, X$).  The first step towards a good description of $\pi$P on nuclear targets is a realistic model of the elementary reaction (on nucleons). The most popular model for this process in neutrino interaction simulations was developed by Rein and Sehgal~\cite{Rein:1980wg}. It assumes that $\pi$P on the nucleon is dominated by baryon resonance excitation, which is described using the relativistic quark model of Feynman, Kissinger
and Ravndal~\cite{Feynman:1971wr} for resonances with invariant masses up to 2~GeV. The model originally neglected final lepton masses. This is a bad approximation at low $Q^2$ but finite mass corrections in kinematics and currents have been recently investigated~\cite{Kuzmin:2004ke,Berger:2007rq,Graczyk:2007xk}. More worrying is the poor description of electron scattering on the proton (see Fig.~2 of Ref.~\cite{Graczyk:2007bc} and Fig.~2 of Ref.~\cite{Leitner:2008fg}) due to the use of unrealistic vector FF. On the other side, the wealth of pion photo- and electro-production data available from several experiments at MIT/Bates, MAMI/Mainz and specially JLab have been used to extract the electromagnetic transition helicity amplitudes~\cite{Drechsel:2007if,Aznauryan:2002gd}. This valuable empirical information should be incorporated to the analysis of neutrino experiments. 

In contrast,  there is almost no information about the axial part of the weak nucleon-to-resonance transition current. PCAC and pion-pole dominance of the pseudoscalar FF can be applied to relate the axial coupling for the dominant contribution at low $Q^2$ to the resonance $\pi N$ decay coupling [off-diagonal Goldberger-Treiman (GT) relation, see for instance Appendix~C of Ref.~\cite{Leitner:2008ue}]. In the few-GeV region, weak $\pi$P is dominated by the excitation of the $\Delta(1232)$ resonance, for which the axial transition current can be cast as 
\begin{align}
-\mathcal{A}^{\mu }_{3/2} &= u_\alpha(p') \left[\frac{C_3^A}{m_N} (g^{\alpha \mu} q \slash  - q^{\alpha} \gamma^{\mu})+
  \frac{C_4^A}{m_N^2} (g^{\alpha \mu} q\cdot p' \right. \nonumber \\
&- q^{\alpha} {p'}^{\mu})  
 \left. + {C_5^A} g^{\alpha \mu}  + \frac{C_6^A}{m_N^2} q^{\alpha} q^{\mu}\right] \gamma_{5}\, u(p) \,. 
\end{align}
For small $Q^2$, only the axial $C_5^A$ FF is relevant and some effort has been devoted to its extraction from ANL and BNL bubble chamber data. In Ref.~\cite{AlvarezRuso:1998hi} $C_5^A(0)$ was extracted from 
the ratio of the inelastic $\nu_\mu \, d \raw \mu^- \, \pi^+ \, p \, n$ and quasielastic $\nu_\mu \, d \raw \mu^- \, p \, p$ $Q^2$ distributions measured at BNL. In the ratio neutrino flux uncertainties largely cancel out. The result $C_5^A(0)=1.22 \pm 0.06$ is compatible with the GT value of 1.2. Assuming a simple dipole parametrization like in Eq.~(3) for  $C_5^A$, Graczyk et al.~\cite{Graczyk:2009qm} obtained $C_5^A(0)=1.19 \pm 0.08$ and $M_{A \Delta} = 0.94 \pm 0.03$~GeV by directly fitting $d \sigma/d Q^2$ for $\nu_\mu \, d \raw \mu^- \, \pi^+ \, p \, n$  ANL and BNL data taking into account normalization uncertainties. Both studies included deuteron effects but neglected nonresonant backgrounds. The nonresonant contribution close to threshold is fully determined by chiral symmetry~\cite{Hernandez:2007qq}. Its inclusion required a considerable reduction of $C^A_5 (0)= 0.867 \pm 0.075$ (with $M_{A \Delta} = 0.985 \pm 0.082$~GeV) to describe the data, but the fit was done to ANL data alone, which are systematically below BNL ones. A recent reanalysis including also low-energy BNL data and deuteron effects finds a higher $C^A_5 (0)=1.00 \pm 0.11$, with $M_{A \Delta} =0.93 \pm 0.07$~GeV~\cite{Hernandez:2010bx}, but still implies a 20~\% reduction of the GT relation. This result is in agreement with  $C^A_5 (0)=0.96$ obtained in the dynamical model for weak  $\pi$P of Ref.~\cite{Sato:2003rq} after a 30~\% renormalization of the constituent quark model prediction by meson clouds.    

When $\pi$P takes place inside the nucleus, the strong-interacting environment leaves a big imprint on the observables. First of all, the elementary amplitude is modified in the medium by the presence of the nuclear mean field and, most importantly, due to the modification of the $\Delta(1232)$ resonance, whose mass gets shifted and width increased because of absorption processes involving one ($\Delta \, N \raw N \, N$) or more nucleons. In addition, the produced pion can be absorbed or scatter with the nucleons with and without charge exchange. In the few-GeV region a large number of states can be excited so that the only feasible way of describing the exclusive final system is with a semiclassical treatment. The most common framework to deal with this is an intranuclear cascade but transport theory has also been used. Reviews and short descriptions of the different MC models applied to $\nu A$ interactions can be found in Refs.~\cite{Gallagher:2009zza,Dytman:2009zza}. Here I shall only refer the main features of nuclear $\pi$P calculations undertaken by theory groups. In the NuWRO event generator~\cite{Juszczak:2005zs} pions are produced via $\Delta(1232)$ excitation. Heavier resonances are only accounted for in the duality-inspired nonresonant background. Pion propagation is accomplished with an intranuclear cascade, with scattering probabilities determined by $\pi N$ vacuum cross sections. Pion absorption is fixed according to pion nuclear absorption data. The model of Ahmad et al.~\cite{Ahmad:2006cy} takes also into account the $\Delta(1232)$ in-medium change only in the production amplitude while the pion cascade uses vacuum cross sections. Finally, the Giessen Boltzmann-Uehling-Uhlenberg (GiBUU) model is a semiclassical transport model in coupled channels successfully applied to photo-, electro- and hadron-nucleus reactions and recently extended to $\nu A$ collisions~\cite{Leitner:2006ww}. Not only the $\Delta(1232)$ but all the baryon resonances with masses up to 2~GeV can be weakly excited and are explicitly propagated. The medium modifications equally affect the production mechanism and secondary interactions.    

\subsection{The incoherent pion production puzzle}

MiniBooNE has measured the ratio $\sigma(\mathrm{CC}\pi^+)/\sigma(\mathrm{CCQE-like})$ on CH$_2$ as a function of $E_\nu$~\cite{AguilarArevalo:2009eb}. There is an uncertainty in the neutrino-energy reconstruction but, on the other side, this observable does not depend on the neutrino flux normalization. The three models described above have been employed to calculate this ratio: see Fig.~8 of Ref.~\cite{Graczyk:2009qm}, Fig.~1 of Ref.~\cite{SajjadAthar:2009rc} and Fig.~2 of Ref.~\cite{Leitner:2008wx}. The comparison shows a good agreement at the lowest energies that gets progressively worse as the energy increases, with NuWRO~\cite{Graczyk:2009qm} getting better agreement while
the GiBUU calculation~\cite{Leitner:2008wx} exhibits the largest discrepancies. Recall however that the three calculations have used $M_A\sim1$~GeV for the CCQE cross section that enters the denominator, which is clearly insufficient to explain MiniBooNE CCQE data as shown in Fig.~\ref{CCQE}. Therefore it turns out all these models underestimate considerably MiniBooNE $\pi$P cross section. Indeed, the comparison of the NC$\pi^0$ pion momentum distribution obtained by GiBUU~\cite{Leitner:2009zz} with the corresponding MiniBooNE data~\cite{AguilarArevalo:2009ww} reveals a good agreement in the shape but a factor of $\sim 2$ discrepancy in the normalization which is a major 
challenge in our understanding of weak $\pi$P.
\bfig[h!]
\bcen
\includegraphics[width=1.03\linewidth]{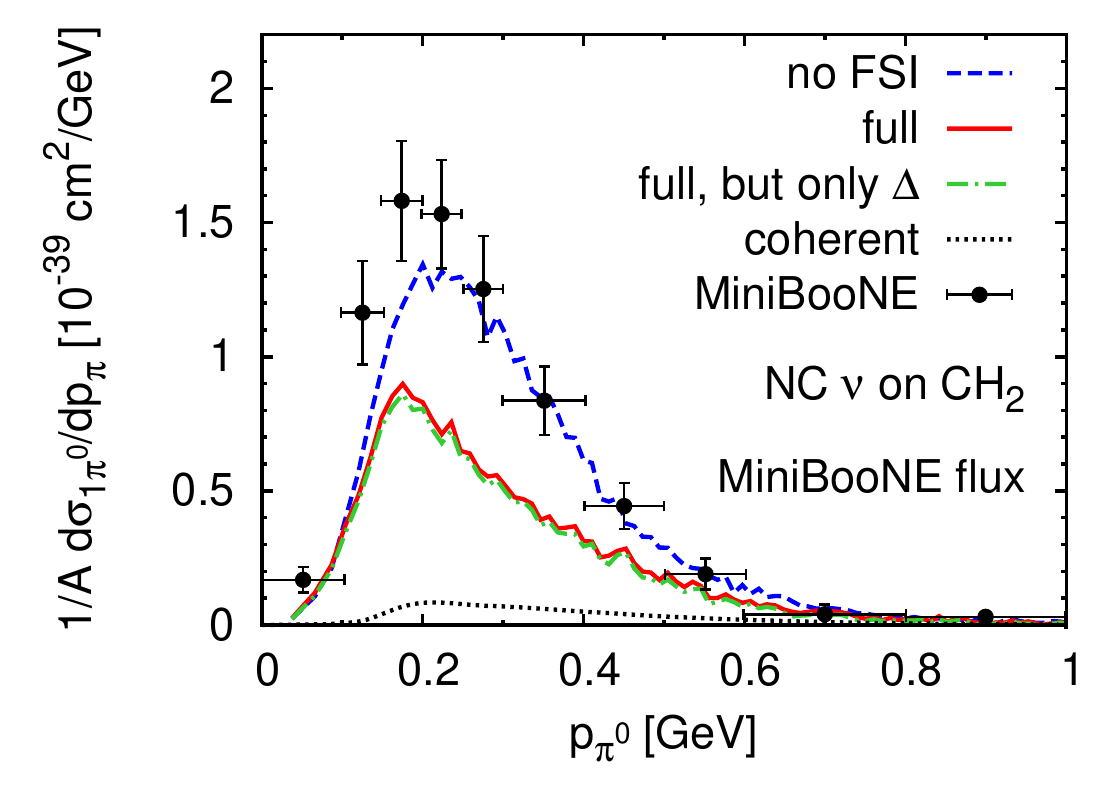}
\caption{Pion momentum distribution for NC$\pi^0$ production on CH$_2$ averaged over the MiniBooNE flux~\cite{AguilarArevalo:2008yp} calculated with the GiBUU transport model~\cite{Leitner:2009zz}. Data are from Ref.~\cite{AguilarArevalo:2009ww}.}
\label{NCpi0}
\ecen
\efig 

\section{Modelling coherent pion production}

Pion production in nuclei is coherent when the nucleus remains in its ground state ($\nu_l \, A \raw l \, \pi \, A$, with $\pi = \pi^+$ in the CC case and $\pi = \pi^0$ in the NC one). It takes place predominantly when the momentum transfered to the the nucleus is small so that the latter is less likely to break. Weak coherent $\pi$P has a very small cross section compared to the incoherent process, but relatively larger than equivalent reactions induced by photons or electrons due to the non vanishing contribution of the axial current at the relevant kinematics~\cite{Amaro:2008hd}. 

The pioneering model of Rein and Sehgal (RS)~\cite{Rein:1982pf} uses PCAC in the $Q^2=0$ limit to relate 
neutrino-induced coherent $\pi$P to pion-nucleus elastic scattering, given in terms of the pion-nucleon cross section. By taking the $Q^2=0$ limit, the RS model neglects important angular dependence at low energies~\cite{Hernandez:2009vm}. This, together with the fact that the description of the pion-nucleus elastic cross section is not realistic (see Fig.~2 of Ref.~\cite{Hernandez:2009vm}), results in cross sections well above the experimental data.
An alternative approach based on PCAC directly uses the experimental pion-nucleus elastic cross section~\cite{Berger:2008xs,Paschos:2009ag}. In this way the treatment of the outgoing pion is improved, but a spurious initial pion distortion, present in pion-nucleus elastic scattering but not in coherent $\pi$P, is introduced. With this method the cross sections are smaller and more compatible with the experiments. As the energy increases, the deficiencies become less relevant and the RS, and PCAC models in general,  becomes realistic. In fact, the recent NOMAD measurement of the NC$\pi^0$ coherent cross section at high (3-100~GeV) energies found $\sigma = [72.6 \pm 8.1(stat) \pm 6.9(syst)] \times 10^{-40}$~cm$^2$ which is consistent with the RS prediction of $78 \times 10^{-40}$~cm$^2$~\cite{Kullenberg:2009pu}. 

Microscopic approaches meant to work in the $\Delta$ region have also been developed. They combine the $\Delta$ excitation picture of weak $\pi$P on the nucleon or the more complete models of Refs.~\cite{Hernandez:2007qq,Sato:2003rq} with the $\Delta$-hole model of pion-nucleus interaction. As the nucleus remains in its ground state, a quantum treatment of pion distortion is feasible by means of the eikonal approximation~\cite{Singh:2006bm}, the Klein-Gordon~\cite{AlvarezRuso:2007tt,Amaro:2008hd} or the Lippmann-Schwinger~\cite{Nakamura:2009iq} equations with realistic pion-nucleus optical potentials. In analogy to the incoherent case, 
it is found that medium effects and pion distortion reduce considerably the cross section and shift the peak to lower pion-momenta (see for example Fig.~2 of Ref.~\cite{Amaro:2008hd}). Nonlocalities in the $\Delta$ propagation were found relevant~\cite{Leitner:2009ph} and have been explicitly taken into account in Ref.~\cite{Nakamura:2009iq} but not in Refs.~\cite{Singh:2006bm,AlvarezRuso:2007tt,AlvarezRuso:2007it,Amaro:2008hd,Hernandez:2010jf}, where they are (at least partially) included via the empirical $\Delta$ mass shift~\cite{Nakamura:2009iq,Hernandez:2010jf}. 

The size of the coherent $\pi$P cross section is very sensitive to the value of $C^A_5 (0)$~\cite{AlvarezRuso:2007it}. This can be easily understood from the fact that at $Q^2 = 0$ (not reachable in the CC case with nonzero lepton masses), where most of the strength of this reaction is concentrated the only FF that contributes is $C^A_5$. Therefore at low energies, to a good approximation, the cross section scales with $[C^A_5 (0)]^2$. For this reason, establishing the degree of violation of the off-diagonal GT relation for the N-$\Delta$ transition as discussed in the previous section is crucial to predict the coherent pion production strength at low energies.

\subsection{The coherent pion production puzzle}

SciBooNE has recently published a measurement of the CC/NC coherent pion production ratio employing a neutrino beam of $\langle E_\nu \rangle \sim 0.8$~GeV~\cite{Kurimoto:2010rc}. They find that 
\be
\frac{\sigma(\mathrm{CCcoh}\pi^+)}{\sigma(\mathrm{NCcoh}\pi^0)} = 0.14^{+0.30}_{-0.28} \,.
\ee 
The theoretical models outlined above predict different cross sections. Different are also the predictions from various MC generators even if all implement the RS model; see Fig.~8 of Ref.~\cite{Boyd:2009zz} for a compilation. However, the ratio is always in the $\mathrm{CC/NC} \sim 1-2$ interval for $E_\nu \sim 0.8$~GeV, which is in strong contradiction with the SciBooNE result. Indeed, neglecting final lepton masses and in the $Q^2=0$ limit this ratio is equal to 2; in a realistic situation one expects only some corrections to this result. For example, in Ref.~\cite{Hernandez:2010jf} the ratio of the flux-averaged cross sections is found to be $1.46 \pm 0.03$, below 2 as one would expect, but far from the SciBooNE value. 

\section{Modelling single photon emission}
 
Like weak $\pi$P, NC single photon production can be incoherent ($\nu \, A \raw \nu \, \gamma \, X$) or coherent ($\nu \, A \raw \nu \, \gamma \, A$). These processes were studied in Ref.~\cite{Hill:2009ek} with effective Lagrangians including nonresonant Compton-like diagrams (Fig.~1 of Ref.~\cite{Hill:2009ek}), $t-$channel contributions from the anomalous vertices with $\pi$, $\rho$ and $\omega$ meson exchange (Fig.~2 of Ref.~\cite{Hill:2009ek}) and $\Delta(1232)$ excitation terms (Fig.~3 of Ref.~\cite{Hill:2009ek}). The cross section on single nucleons and the coherent nuclear one were computed. The $\Delta(1232)$ contribution was found to be the largest one in both coherent and incoherent cases. The $t-$channel part is dominated by $\omega$ exchange. It was found that MiniBooNE's strategy of normalizing $\Delta \raw N \, \gamma$ events to the measured $\Delta \raw N \, \pi$ rate is a reasonable first approximation for the incoherent reaction but not for the coherent one. Within the uncertainties, the model seems capable of providing enough photons to cover the low-energy excess found by MiniBooNE~\cite{Hill:2010zy} but a precise account of nuclear effects is necessary.

\section{Concluding remarks}

New neutrino interaction data with high statistics, accompanied by a better understanding of the neutrino fluxes are becoming available from several experiments. The comparison with theory for some of the reaction channels that are relevant for oscillations studies and interesting from the perspective of hadronic and nuclear physics reveals discrepancies that await explanation: fitting the new results with the available parameters is a dangerous strategy in the long term. The general tendency is that theory underestimates data but some flux-normalization independent quantities like the $\sigma(\mathrm{CC}\pi^+)/\sigma(\mathrm{CCQE-like})$ ratio are also not well described by state-of-the-art models. CC coherent $\pi^+$ production turned out to be elusive even if all theoretical models predict its cross section to be larger than the NC coherent $\pi^0$ one. Future data and ongoing theoretical work shall be quite helpful to clarify the situation. In order to achieve the precision goals in neutrino oscillation measurements and to reliably extract information about the axial properties of the nucleon and baryon resonances it is crucial that the current theoretical developments are implemented in the event generators used in the experimental data analysis.       

\section*{Acknowledgements} 

I thank Costas Andreopoulos and Tina Leitner for providing Figs.~\ref{genie} and \ref{NCpi0}, and the organizers for the opportunity to present this review.

\bibliographystyle{elsarticle-num}
\bibliography{Neutrino_mod}

\begin{thebibliography}{10}
\expandafter\ifx\csname url\endcsname\relax
  \def\url#1{\texttt{#1}}\fi
\expandafter\ifx\csname urlprefix\endcsname\relax\def\urlprefix{URL }\fi
\expandafter\ifx\csname href\endcsname\relax
  \def\href#1#2{#2} \def\path#1{#1}\fi

\bibitem{Andreopoulos:2009rq}
C.~Andreopoulos, et~al., Nucl. Instrum. Meth. A614 (2010) 87--104.

\bibitem{Leitner:2010kp}
T.~Leitner, U.~Mosel, Phys. Rev. C81 (2010) 064614.

\bibitem{AguilarArevalo:2007it}
A.~A. Aguilar-Arevalo, et~al., Phys. Rev. Lett. 98 (2007) 231801.

\bibitem{AguilarArevalo:2010zc}
A.~A. Aguilar-Arevalo, et~al., Phys. Rev. D81 (2010) 092005.

\bibitem{AguilarArevalo:2010cx}
A.~A. Aguilar-Arevalo, et~al.\href {http://arxiv.org/abs/1007.4730} {\path{,
  arXiv:1007.4730}}.

\bibitem{Lyubushkin:2008pe}
V.~Lyubushkin, et~al., Eur. Phys. J. C63 (2009) 355--381.

\bibitem{AguilarArevalo:2010bm}
A.~A. Aguilar-Arevalo, et~al.\href {http://arxiv.org/abs/1011.3572} {\path{,
  arXiv:1011.3572}}.

\bibitem{AguilarArevalo:2010xt}
A.~A. Aguilar-Arevalo, et~al.\href {http://arxiv.org/abs/1010.3264} {\path{,
  arXiv:1010.3264}}.

\bibitem{AguilarArevalo:2009ww}
A.~A. Aguilar-Arevalo, et~al., Phys. Rev. D81 (2010) 013005.

\bibitem{Hiraide:2008eu}
K.~Hiraide, et~al., Phys. Rev. D78 (2008) 112004.

\bibitem{Kurimoto:2010rc}
Y.~Kurimoto, et~al., Phys. Rev. D81 (2010) 111102.

\bibitem{Kullenberg:2009pu}
C.~T. Kullenberg, et~al., Phys. Lett. B682 (2009) 177--184.

\bibitem{Bodek:2007ym}
A.~Bodek, S.~Avvakumov, R.~Bradford, H.~S. Budd, Eur. Phys. J. C53 (2008)
  349--354.

\bibitem{Liesenfeld:1999mv}
A.~Liesenfeld, et~al., Phys. Lett. B468 (1999) 20.

\bibitem{Bernard:1992ys}
V.~Bernard, N.~Kaiser, U.~G. Meissner, Phys. Rev. Lett. 69 (1992) 1877--1879.

\bibitem{Smith:1972xh}
R.~A. Smith, E.~J. Moniz, Nucl. Phys. B43 (1972) 605.

\bibitem{Benhar:2005dj}
O.~Benhar, N.~Farina, H.~Nakamura, M.~Sakuda, R.~Seki, Phys. Rev. D72 (2005)
  053005.

\bibitem{Ankowski:2007uy}
A.~M. Ankowski, J.~T. Sobczyk, Phys. Rev. C77 (2008) 044311.

\bibitem{Maieron:2003df}
C.~Maieron, M.~C. Martinez, J.~A. Caballero, J.~M. Udias, Phys. Rev. C68 (2003)
  048501.

\bibitem{Meucci:2004ip}
A.~Meucci, C.~Giusti, F.~D. Pacati, Nucl. Phys. A744 (2004) 307--322.

\bibitem{Leitner:2008ue}
T.~Leitner, O.~Buss, L.~Alvarez-Ruso, U.~Mosel, Phys. Rev. C79 (2009) 034601.

\bibitem{Nieves:2004wx}
J.~Nieves, J.~E. Amaro, M.~Valverde, Phys. Rev. C70 (2004) 055503.

\bibitem{Singh:1992dc}
S.~K. Singh, E.~Oset, Nucl. Phys. A542 (1992) 587--615.

\bibitem{Martini:2009uj}
M.~Martini, M.~Ericson, G.~Chanfray, J.~Marteau, Phys. Rev. C80 (2009) 065501.

\bibitem{Boyd:2009zz}
S.~Boyd, S.~Dytman, E.~Hernandez, J.~Sobczyk, R.~Tacik, AIP Conf. Proc. 1189
  (2009) 60--73.

\bibitem{SajjadAthar:2009rd}
M.~Sajjad~Athar, S.~Chauhan, S.~K. Singh, Eur. Phys. J. A43 (2010) 209--227.

\bibitem{Benhar:2006nr}
O.~Benhar, D.~Meloni, Nucl. Phys. A789 (2007) 379--402.

\bibitem{Leitner:2006ww}
T.~Leitner, L.~Alvarez-Ruso, U.~Mosel, Phys. Rev. C73 (2006) 065502.

\bibitem{AguilarArevalo:2008yp}
A.~A. Aguilar-Arevalo, et~al., Phys. Rev. D79 (2009) 072002.

\bibitem{Butkevich:2010cr}
A.~V. Butkevich\href {http://arxiv.org/abs/1006.1595} {\path{,
  arXiv:1006.1595}}.

\bibitem{Benhar:2010nx}
O.~Benhar, P.~Coletti, D.~Meloni, Phys. Rev. Lett. 105 (2010) 132301.

\bibitem{Martini:2010ex}
M.~Martini, M.~Ericson, G.~Chanfray, J.~Marteau, Phys. Rev. C81 (2010) 045502.

\bibitem{Amaro:2010sd}
J.~E. Amaro, M.~B. Barbaro, J.~A. Caballero, T.~W. Donnelly, C.~F.
  Williamson\href {http://arxiv.org/abs/1010.1708} {\path{, arXiv:1010.1708}}.

\bibitem{Rein:1980wg}
D.~Rein, L.~M. Sehgal, Ann. Phys. 133 (1981) 79.

\bibitem{Feynman:1971wr}
R.~P. Feynman, M.~Kislinger, F.~Ravndal, Phys. Rev. D3 (1971) 2706--2732.

\bibitem{Kuzmin:2004ke}
K.~S. Kuzmin, V.~V. Lyubushkin, V.~A. Naumov, Mod. Phys. Lett. A19 (2004)
  2919--2928.

\bibitem{Berger:2007rq}
C.~Berger, L.~M. Sehgal, Phys. Rev. D76 (2007) 113004.

\bibitem{Graczyk:2007xk}
K.~M. Graczyk, J.~T. Sobczyk, Phys. Rev. D77 (2008) 053003.

\bibitem{Graczyk:2007bc}
K.~M. Graczyk, J.~T. Sobczyk, Phys. Rev. D77 (2008) 053001.

\bibitem{Leitner:2008fg}
T.~Leitner, O.~Buss, U.~Mosel, L.~Alvarez-Ruso, PoS NUFACT08 (2008) 009.

\bibitem{Drechsel:2007if}
D.~Drechsel, S.~S. Kamalov, L.~Tiator, Eur. Phys. J. A34 (2007) 69--97.

\bibitem{Aznauryan:2002gd}
I.~G. Aznauryan, Phys. Rev. C67 (2003) 015209.

\bibitem{AlvarezRuso:1998hi}
L.~Alvarez-Ruso, S.~K. Singh, M.~J. Vicente~Vacas, Phys. Rev. C59 (1999)
  3386--3392.

\bibitem{Graczyk:2009qm}
K.~M. Graczyk, D.~Kielczewska, P.~Przewlocki, J.~T. Sobczyk, Phys. Rev. D80
  (2009) 093001.

\bibitem{Hernandez:2007qq}
E.~Hernandez, J.~Nieves, M.~Valverde, Phys. Rev. D76 (2007) 033005.

\bibitem{Hernandez:2010bx}
E.~Hernandez, J.~Nieves, M.~Valverde, M.~J. Vicente~Vacas, Phys. Rev. D81
  (2010) 085046.

\bibitem{Sato:2003rq}
T.~Sato, D.~Uno, T.~S.~H. Lee, Phys. Rev. C67 (2003) 065201.

\bibitem{Gallagher:2009zza}
H.~R. Gallagher, AIP Conf. Proc. 1189 (2009) 35--42.

\bibitem{Dytman:2009zza}
S.~Dytman, AIP Conf. Proc. 1189 (2009) 51--59.

\bibitem{Juszczak:2005zs}
C.~Juszczak, J.~A. Nowak, J.~T. Sobczyk, Nucl. Phys. Proc. Suppl. 159 (2006)
  211--216.

\bibitem{Ahmad:2006cy}
S.~Ahmad, M.~Sajjad~Athar, S.~K. Singh, Phys. Rev. D74 (2006) 073008.

\bibitem{AguilarArevalo:2009eb}
A.~A. Aguilar-Arevalo, et~al., Phys. Rev. Lett. 103 (2009) 081801.

\bibitem{SajjadAthar:2009rc}
M.~Sajjad~Athar, S.~Chauhan, S.~K. Singh, J. Phys. G37 (2010) 015005.

\bibitem{Leitner:2008wx}
T.~Leitner, O.~Buss, U.~Mosel, L.~Alvarez-Ruso, Phys. Rev. C79 (2009) 038501.

\bibitem{Leitner:2009zz}
T.~J. Leitner, Ph.D. thesis, Giessen University (2009).

\bibitem{Amaro:2008hd}
J.~E. Amaro, E.~Hernandez, J.~Nieves, M.~Valverde, Phys. Rev. D79 (2009)
  013002.

\bibitem{Rein:1982pf}
D.~Rein, L.~M. Sehgal, Nucl. Phys. B223 (1983) 29.

\bibitem{Hernandez:2009vm}
E.~Hernandez, J.~Nieves, M.~J. Vicente-Vacas, Phys. Rev. D80 (2009) 013003.

\bibitem{Berger:2008xs}
C.~Berger, L.~M. Sehgal, Phys. Rev. D79 (2009) 053003.

\bibitem{Paschos:2009ag}
E.~A. Paschos, D.~Schalla, Phys. Rev. D80 (2009) 033005.

\bibitem{Singh:2006bm}
S.~K. Singh, M.~Sajjad~Athar, S.~Ahmad, Phys. Rev. Lett. 96 (2006) 241801.

\bibitem{AlvarezRuso:2007tt}
L.~Alvarez-Ruso, L.~S. Geng, S.~Hirenzaki, M.~J. Vicente~Vacas, Phys. Rev. C75
  (2007) 055501.

\bibitem{Nakamura:2009iq}
S.~X. Nakamura, T.~Sato, T.~S.~H. Lee, B.~Szczerbinska, K.~Kubodera, Phys. Rev.
  C81 (2010) 035502.

\bibitem{Leitner:2009ph}
T.~Leitner, U.~Mosel, S.~Winkelmann, Phys. Rev. C79 (2009) 057601.

\bibitem{AlvarezRuso:2007it}
L.~Alvarez-Ruso, L.~S. Geng, M.~J. Vicente~Vacas, Phys. Rev. C76 (2007) 068501.

\bibitem{Hernandez:2010jf}
E.~Hernandez, J.~Nieves, M.~Valverde, Phys. Rev. D82 (2010) 077303.

\bibitem{Hill:2009ek}
R.~J. Hill, Phys. Rev. D81 (2010) 013008.

\bibitem{Hill:2010zy}
R.~J. Hill\href {http://arxiv.org/abs/1002.4215} {\path{, arXiv:1002.4215}}.

\end{thebibliography}

\end{document}